\documentclass[aps,showpacs,preprintnumbers,amsmath, amssymb]{revtex4}

\usepackage{float}
\usepackage{graphics,epsfig}
\usepackage{graphicx}
\usepackage{dcolumn}
\usepackage{bm}

\begin{document}
\baselineskip=0.8 cm
\title{{\bf Analytical study on holographic superconductors with backreactions}}

\author{Qiyuan Pan$^{1,2}$\footnote{panqiyuan@126.com}, Jiliang Jing$^{1,2}$\footnote{jljing@hunnu.edu.cn}, Bin Wang$^{3}$\footnote{wang\_b@sjtu.edu.cn} and Songbai Chen$^{1,2}$\footnote{csb3752@163.com}}
\affiliation{$^{1}$Institute of Physics and Department of Physics,
Hunan Normal University, Changsha, Hunan 410081, China}
\affiliation{$^{2}$ Key Laboratory of Low Dimensional Quantum
Structures and Quantum Control of Ministry of Education, Hunan
Normal University, Changsha, Hunan 410081, China}
\affiliation{$^{3}$ INPAC and Department of Physics, Shanghai Jiao
Tong University, Shanghai 200240, China}

\vspace*{0.2cm}
\begin{abstract}
\baselineskip=0.6 cm
\begin{center}
{\bf Abstract}
\end{center}

We employ the variational method for the Sturm-Liouville
eigenvalue problem to analytically investigate the properties of the
holographic superconductors. We find that the analytic method is
still powerful when the backreaction is turned on. Reducing step
size in the iterative procedure, we observe that the consistency of
results between the analytic and numerical computations can be
further improved. The obtained analytic result can be used to back
up the numerical computations in the holographic superconductor in
the fully backreacted spacetime.

\end{abstract}

\maketitle
\newpage
\vspace*{0.2cm}

\section{Introduction}

As a powerful tool to analyse strongly coupled quantum field
theories, the anti-de Sitter/conformal field theories (AdS/CFT)
correspondence \cite{Maldacena,Witten,Gubser1998} states that a
$d$-dimensional weakly coupled dual gravitational description in the
bulk is equivalent to a $(d-1)$-dimensional strongly coupled
conformal field theory on the boundary. In recent years, this
principle has been used to provide some meaningful theoretical
insights in order to understand the physics of high $T_{c}$
superconductors from the gravitational dual
\cite{HartnollPRL101,HartnollJHEP12}. It was found that the
spontaneous $U(1)$ symmetry breaking by bulk black holes can be used
to construct gravitational duals of the transition from normal state
to superconducting state in the boundary theory, which exhibits the
behavior of the superconductor \cite{GubserPRD78}. Due to the
potential applications to the condensed matter physics, there have
been a lot of works studying various gravity models with the
property of the so-called holographic superconductor (for reviews,
see Refs. \cite{HartnollRev,HerzogRev,HorowitzRev} and references
therein).

In most cases, the studies on the holographic superconductors focus
on the probe approximation where the backreaction of matter fields
on the spacetime metric is neglected. When taking the backreaction
into account, it was found that even the uncharged scalar field can
form a condensate in the $(2+1)$-dimensional holographic
superconductor model \cite{HartnollJHEP12}. Furthermore, in the
p-wave holographic dual models, it was argued that the phase
transition that leads to the formation of vector hair changes from
the second order to the first order when the gravitational coupling
is large enough \cite{Ammon2010,CaiBackreaction}. There have been
accumulated interest to study the holographic superconductor away
from the probe limit
\cite{Gubser-Nellore,Aprile-Russo,Brihaye,Liu-Sun,BarclayGregory,Siani,
Barclay2011,PanWangBR,LiuWangBTZ,GregoryRev,LiuBIBR,HGST2}. Recently, the
effect of the backreaction has been investigated between holographic
insulator and superconductor
\cite{Horowitz-Soliton,PengWangPLB,Akhavan-Soliton,brihaye-Soliton,PengBR}.

Almost all works on the holographic dual models away from the probe
limit were based on numerical computations. In order to back up
numerical results and gain more insights in the effect of the
backreaction,  a fully analytic study is called for. Refining the
analytic matching method developed in \cite{Gregory,Pan-Wang}, Kanno
calculated the critical temperature of ($3+1$)-dimensional
holographic superconductors in Einstein-Gauss-Bonnet gravity with
backreaction and found that the backreaction makes condensation
harder \cite{KannoCQG}. This analytic approach has been extended to
derive the critical magnetic field in holographic superconductors
with backreaction \cite{Ge2011}. However, the analytic matching
method can keep valid only when the matching point is chosen within
an appropriate range in higher dimensions ($d>6$) \cite{Pan-Wang}. Moreover, when
the scalar mass is zero in the Gauss-Bonnet holographic
superconductors, the curvature correction term does not contribute
to the analytic approximation, which leads the analytic procedure to
break down \cite{Pan-Wang}. Recently, Siopsis and Therrien developed
a new analytic method. They extended the variational method for the
Sturm-Liouville (S-L) eigenvalue problem to analytically calculate
the critical exponent near the critical temperature and found that
the analytical results obtained by this way are in good agreement
with the numerical findings \cite{Siopsis}. Considering the
effectiveness and accuracy of the S-L method, many authors have used
it to analytically investigate the properties of holographic
superconductors in AdS black hole backgrounds
\cite{SiopsisBF,ZengSL,Li-Cai-Zhang,JingJHEP,MomeniSL,HGST1,SDSL1,SDSL2}
and soliton backgrounds
\cite{CaiLiZhang,CaiSoliton,PanJingWang,LeeSL}. But these attempts
were limited in the probe limit. It was argued that the S-L method
is more effective for the analytic study of the condensation than
the matching method [42]. It is of interest to examine whether the
S-L method is still valid to explore the holographic
superconductivity when the backreaction is turned on. This is not
trivial since the analytic study can help to confirm the numerical
result. Furthermore it can clearly disclose the influence of the
role of the backreaction in the condensation. In this work, we will
generalize the variational method for the S-L eigenvalue problem to
study holographic superconductor away from the probe limit.

The organization of the work is as follows. In Sec. II, we will
introduce the holographic superconductor models with backreactions
in the $d$-dimensional AdS black hole background. In Sec. III we will
give an analytical investigation of the holographic superconductors
by using the S-L method. We will conclude in the last section of our
main results.

\section{Holographic superconductor models with backreactions}

We begin with the general action describing a charged, complex
scalar field in the $d$-dimensional Einstein-Maxwell action with
negative cosmological constant
\begin{eqnarray}\label{System}
S=\int
d^{d}x\sqrt{-g}\left[\frac{1}{2\kappa^{2}}\left(R-2\Lambda\right)
-\frac{1}{4}F_{\mu\nu}F^{\mu\nu}-|\nabla\psi-iqA\psi|^{2}
-m^2|\psi|^2\right],
\end{eqnarray}
where $\kappa^{2}=8\pi G_{d}$ is the $d$-dimensional gravitational
constant, $\Lambda=-(d-1)(d-2)/(2L^{2})$ is the cosmological
constant, $A$ and $\psi$ represent the gauge field and a scalar
field with charge $q$ respectively. Since we are interested in
including the backreaction, we will take the metric ansatz for the
$d$-dimensional planar black hole
\begin{eqnarray}\label{BH metric}
ds^2=-f(r)e^{-\chi(r)}dt^{2}+\frac{dr^2}{f(r)}+r^{2}h_{ij}dx^{i}dx^{j},
\end{eqnarray}
where $f$ and $\chi$ are functions of $r$ only, $h_{ij}dx^{i}dx^{j}$
denotes the line element of a ($d-2$)-dimensional hypersurface with
the curvature $k=0$. The Hawking temperature of this black hole,
which will be interpreted as the temperature of the CFT, is given by
\begin{eqnarray}\label{Hawking temperature}
T_{H}=\frac{f^{\prime}(r_{+})e^{-\chi(r_{+})/2}}{4\pi},
\end{eqnarray}
where the prime denotes a derivative with respect to $r$.  $r_{+}$
is the black hole horizon determined by $f(r_{+})=0$.

We consider the electromagnetic field and the scalar field in the
forms
\begin{eqnarray}
A=\phi(r)dt,~~\psi=\psi(r),
\end{eqnarray}
where without loss of generality $\psi(r)$ can be taken to be real.
Thus, from the variation of the action with respect to the matter
and metric we obtain the equations of motion
\begin{eqnarray}
\chi^{\prime}+\frac{4\kappa^{2}r}{d-2}\left(\psi^{\prime
2}+\frac{q^{2}e^{\chi}\phi^{2}\psi^{2}}{f^{2}}\right)=0,\label{chi}
\end{eqnarray}
\begin{eqnarray}\label{fr}
f^{\prime}-\left[\frac{(d-1)r}{L^{2}}-\frac{(d-3)f}{r}\right]+\frac{2\kappa^{2}r}{d-2}
\left[m^{2}\psi^{2}+\frac{1}{2}e^{\chi}\phi^{\prime
2}+f\left(\psi^{\prime
2}+\frac{q^{2}e^{\chi}\phi^{2}\psi^{2}}{f^{2}}\right)\right]=0,
\end{eqnarray}
\begin{eqnarray}\label{phi}
\phi^{\prime\prime}+\left(\frac{d-2}{r}+\frac{\chi^{\prime}}{2}\right)\phi^\prime-\frac{2q^{2}\psi^{2}}{f}\phi=0,
\end{eqnarray}
\begin{eqnarray}\label{psi}
\psi^{\prime\prime}+\left(\frac{d-2}{r}-\frac{\chi^{\prime}}{2}+
\frac{f^\prime}{f}\right)\psi^\prime
-\frac{m^2}{f}\psi+\frac{q^{2}e^{\chi}\phi^2}{f^2}\psi=0.
\end{eqnarray}
It should be noted that the transformation $\tilde{\phi}=\phi/q$ and
$\tilde{\psi}=\psi/q$ in the action (\ref{System}) does not change
the form of the Maxwell and the scalar equations, but the
gravitational coupling in the Einstein equation changes
$\kappa^{2}\rightarrow\kappa^{2}/q^{2}$. Thus, the probe limit is
equivalent to letting $q\rightarrow\infty$. Without loss of
generality, we can set $q=1$ and keep $\kappa^{2}$ finite when we
take the backreaction into account \cite{BarclayGregory,
Barclay2011,GregoryRev,KannoCQG}.

For the normal phase, $\psi(r)=0$, we find that $\chi$ is a constant
and the analytic solutions to Eqs. (\ref{fr}) and (\ref{phi}) lead
to the
 AdS Reissner-Nordstr\"{o}m
black holes with the metric coefficient
\begin{eqnarray}
f=\frac{r^{2}}{L^{2}}-\frac{1}{r^{d-3}}\left[\frac{r^{d-1}_{+}}{L^{2}}+\frac{(d-3)\kappa^{2}\rho^{2}}{(d-2)r^{d-3}_{+}}\right]
+\frac{(d-3)\kappa^{2}\rho^{2}}{(d-2)r^{2d-6}}\,,\hspace{0.5cm}
\phi=\mu-\frac{\rho}{r^{d-3}}\,,
\end{eqnarray}
where $\mu$ and $\rho$ are interpreted as the chemical potential and
charge density in the dual field theory respectively. When
$\kappa=0$, the metric coefficient $f$ goes back to the case of the
Schwarzschild AdS black hole.

In order to get the solutions in superconducting phase, where
$\psi(r)\neq0$, we have to count on the appropriate boundary
conditions. At the horizon $r_{+}$, the metric functions $\chi$ and
$f$ satisfy
\begin{eqnarray}
&&\chi^{\prime}(r_{+})=-\frac{4\kappa^{2}r_{+}}{d-2}\left[\psi^{\prime}(r_{+})^{2}+\frac{e^{\chi(r_{+})}\phi^{\prime}(r_{+})^{2}\psi(r_{+})^{2}}{f^{\prime}(r_{+})^{2}}\right],
\nonumber\\
&&f^{\prime}(r_{+})=\frac{(d-1)r_{+}}{L^{2}}-\frac{2\kappa^{2}
r_{+}}{d-2}\left[m^{2}\psi(r_{+})^{2}+\frac{1}{2}e^{\chi(r_{+})}\phi^{\prime}(r_{+})^{2}\right],
\end{eqnarray}
and the regularity condition  gives the boundary conditions
\begin{eqnarray}
\phi(r_{+})=0\,,\hspace{0.5cm}
\psi(r_{+})=\frac{f^\prime(r_{+})\psi^\prime(r_{+})}{m^{2}}.
\end{eqnarray}
At the asymptotic AdS boundary ($r\rightarrow\infty$), the
asymptotic behaviors of the solutions are
\begin{eqnarray}
\chi\rightarrow0\,,\hspace{0.5cm}
f\sim\frac{r^{2}}{L^{2}}\,,\hspace{0.5cm}
\phi\sim\mu-\frac{\rho}{r^{d-3}}\,,\hspace{0.5cm}
\psi\sim\frac{\psi_{-}}{r^{\Delta_{-}}}+\frac{\psi_{+}}{r^{\Delta_{+}}}\,,
\label{infinity}
\end{eqnarray}
where the exponent $\Delta_\pm$ is defined by
$[(d-1)\pm\sqrt{(d-1)^{2}+4m^{2}}]/2$. Notice that, provided $\Delta_{-}$ is larger than the unitarity
bound, both $\psi_{-}$ and $\psi_{+}$ can be normalizable and they can be used to define
operators on the dual field theory, $\psi_{-}=<\mathcal{O}_{-}>$,
$\psi_{+}=<\mathcal{O}_{+}>$, respectively
\cite{HartnollPRL101,HartnollJHEP12}. For simplicity, we will scale
$L=1$ in the following calculation.

\section{Analytical investigation of the holographic superconductors}

Here we will apply the S-L method \cite{Siopsis} to analytically
investigate the properties of the s-wave holographic superconductor
phase transition with backreactions. We will derive the relation
between the critical temperature $T_{c}$ and charge density $\rho$
near the phase transition point and  examine the effect of the
backreaction.

Introducing a new variable $z=r_{+}/r$, we can rewrite the Einstein,
Maxwell and the scalar equations into
\begin{eqnarray}
\chi^{\prime}-\frac{4\kappa^{2}}{d-2}\left(z\psi^{\prime
2}+\frac{r^{2}_{+}}{z^{3}f^{2}}e^{\chi}\phi^{2}\psi^{2}\right)=0,\label{chiz}
\end{eqnarray}
\begin{eqnarray}\label{frz}
f^{\prime}-\frac{(d-3)f}{z}+\frac{(d-1)r^{2}_{+}}{L^{2}z^{3}}-\frac{2\kappa^{2}r^{2}_{+}}{(d-2)z^{3}}
\left[m^{2}\psi^{2}+\frac{z^{4}}{2r^{2}_{+}}e^{\chi}\phi^{\prime
2}+f\left(\frac{z^{4}}{r^{2}_{+}}\psi^{\prime
2}+\frac{1}{f^{2}}e^{\chi}\phi^{2}\psi^{2}\right)\right]=0,
\end{eqnarray}
\begin{eqnarray}\label{phiz}
\phi^{\prime\prime}+\left(\frac{\chi^{\prime}}{2}-\frac{d-4}{z}\right)\phi^\prime-\frac{2r^{2}_{+}\psi^{2}}{z^{4}f}\phi=0,
\end{eqnarray}
\begin{eqnarray}\label{psiz}
\psi^{\prime\prime}-\left(\frac{\chi^{\prime}}{2}+\frac{d-4}{z}-
\frac{f^\prime}{f}\right)\psi^\prime
-\frac{r^{2}_{+}}{z^{4}}\left(\frac{m^2}{f}-\frac{e^{\chi}\phi^2}{f^2}\right)\psi=0,
\end{eqnarray}
where the prime now denotes the derivative with respect to  $z$.

Since the value of the scalar operator $<\mathcal{O}_{+}>$ (or
$<\mathcal{O}_{-}>$) is small near the critical point, we can
introduce it as an expansion parameter
\begin{eqnarray}
\epsilon\equiv<\mathcal{O}_{i}>,
\end{eqnarray}
with $i=+$ or $i=-$. Note that we are interested in solutions where
$\psi$ is small, therefore from Eqs. (\ref{phiz}) and (\ref{psiz})
we can expand the scalar field $\psi$ and the gauge field $\phi$ as
\cite{KannoCQG,Ge2011,Herzog2010}
\begin{eqnarray}
&&\psi=\epsilon\psi_{1}+\epsilon^{3}\psi_{3}+\epsilon^{5}\psi_{5}+\cdot\cdot\cdot,\nonumber\\
&&\phi=\phi_{0}+\epsilon^{2}\phi_{2}+\epsilon^{4}\phi_{4}+\cdot\cdot\cdot,
\end{eqnarray}
where $\epsilon\ll1$. The metric function $f(z)$ and $\chi(z)$ can
be expanded around the Reissner-Nordstr\"{o}m AdS spacetime
\begin{eqnarray}
&&f=f_{0}+\epsilon^{2}f_{2}+\epsilon^{4}f_{4}+\cdot\cdot\cdot,\nonumber\\
&&\chi=\epsilon^{2}\chi_{2}+\epsilon^{4}\chi_{4}+\cdot\cdot\cdot.
\end{eqnarray}
For the chemical potential $\mu$, we will allow it to be corrected
order by order \cite{Herzog2010}
\begin{eqnarray}
\mu=\mu_{0}+\epsilon^{2}\delta\mu_{2}+\cdot\cdot\cdot,
\end{eqnarray}
where $\delta\mu_{2}>0$. Thus, near the phase transition, we find a
result for the order parameter as a function of the chemical
potential
\begin{eqnarray}
\epsilon\approx\left(\frac{\mu-\mu_{0}}{\delta\mu_{2}}\right)^{1/2},
\end{eqnarray}
whose critical exponent $\beta=1/2$ is the universal result from the
Ginzburg-Landau mean field theory of phase transitions. Obviously,
the order parameter becomes zero and phase transition can happen if
$\mu\rightarrow\mu_{0}$, which shows that the critical value of
$\mu$ is $\mu_{c}=\mu_{0}$.

At the zeroth order, we can get the solution $\phi_{0}$ from Eq.
(\ref{phiz}), i.e., the electromagnetic field behaves like
$\phi_{0}(z)=\mu_{0}(1-z^{d-3})$, which gives a relation
$\mu_{0}=\rho/r^{d-3}_{+}$. At the critical point $\mu_{c}$, we can
find $\mu_{0}=\mu_{c}=\rho/r^{d-3}_{+c}$, where $r_{+c}$ is the
radius of the horizon at the critical point. In order to use the
analytical S-L method \cite{Siopsis}, we will set
\begin{eqnarray}
\phi_{0}(z)=\lambda r_{+c}(1-z^{d-3}), \label{Phi-critical}
\end{eqnarray}
with $\lambda=\rho/r^{d-2}_{+c}$. Inserting this solution into Eq.
(\ref{frz}), we obtain the metric function
\begin{eqnarray}
f_{0}(z)=r^{2}_{+}g(z)=r^{2}_{+}\left[\frac{1}{L^{2}z^{2}}-\frac{z^{d-3}}{L^{2}}-\frac{(d-3)\kappa^{2}\lambda^{2}}{d-2}z^{d-3}(1-z^{d-3})\right],
\label{f-critical}
\end{eqnarray}
where we define a new function $g(z)$ for simplicity in the
following calculation.

At the first order, the asymptotic AdS boundary conditions
($z\rightarrow0$) for $\psi$ can be expressed as
\begin{eqnarray}
\psi_{1}\sim\frac{\psi_{-}}{r_{+}^{\Delta_{-}}}z^{\Delta_{-}}+\frac{\psi_{+}}{r_{+}^{\Delta_{+}}}z^{\Delta_{+}}\,.
\end{eqnarray}
So we introduce a trial function $F(z)$ near the boundary $z=0$
\cite{Siopsis}
\begin{eqnarray}\label{phiFz}
\psi_{1}(z)\sim \frac{\langle{\cal
O}_{i}\rangle}{r^{\Delta_{i}}_{+}} z^{\Delta_{i}}F(z),
\end{eqnarray}
where we have imposed the boundary condition $F(0)=1$ and $F'(0)=0$.
Substituting Eq. (\ref{phiFz}) into Eq. (\ref{psiz}), we obtain the
equation of motion for $F(z)$
\begin{eqnarray}\label{Fzmotion}
F^{\prime\prime}+\left[\frac{2(\Delta_i+2)-d}{z}+\frac{g'}{g}\right]
F^{\prime}+\left[\frac{\Delta_{i}}{z}\left(\frac{\Delta_{i}+3-d}{z}+\frac{g'}{g}\right)+\frac{\lambda^{2}(1-z^{d-3})^{2}}{z^{4}g^{2}}
-\frac{m^{2}}{z^{4}g}\right]F=0.
\end{eqnarray}
In order to simplify the following calculation, we will express the
backreacting parameter $\kappa$ as
\begin{eqnarray}\label{Kappa}
\kappa_{n}=n\Delta\kappa,~~~n=0,1,2,\cdot\cdot\cdot,
\end{eqnarray}
where $\Delta\kappa=\kappa_{n+1}-\kappa_{n}$ is the step size of our
iterative procedure. Considering the fact that
$\kappa^{2}\lambda^{2}=\kappa_{n}^{2}\lambda^{2}=\kappa_{n}^{2}(\lambda^{2}|_{\kappa_{n-1}})+0[(\Delta\kappa)^{4}]$
(note that we have set $\kappa_{-1}=0$ and
$\lambda^{2}|_{\kappa_{-1}}=0$), we will use the following form of
$g(z)$ in our discussion
\begin{eqnarray}\label{gzNew}
g(z)\approx\frac{1}{L^{2}z^{2}}-\frac{z^{d-3}}{L^{2}}-\frac{(d-3)\kappa_{n}^{2}(\lambda^{2}|_{\kappa_{n-1}})}{d-2}z^{d-3}(1-z^{d-3}),
\end{eqnarray}
where $\lambda^{2}|_{\kappa_{n-1}}$ is the value of $\lambda^{2}$
for $\kappa_{n-1}$. After defining a function which obeys
\begin{eqnarray}\label{TzNew}
T(z)=z^{2\Delta_{i}+1}[(d-2)(z^{1-d}-1)-(d-3)L^{2}\kappa_{n}^{2}(\lambda^{2}|_{\kappa_{n-1}})(1-z^{d-3})],
\end{eqnarray}
we can convert Eq. (\ref{Fzmotion}) to be
\begin{eqnarray}\label{TFzmotion}
(TF^{\prime})^{\prime}+T\left[\frac{\Delta_{i}}{z}\left(\frac{\Delta_{i}+3-d}{z}+\frac{g'}{g}\right)+\frac{\lambda^{2}(1-z^{d-3})^{2}}{z^{4}g^{2}}
-\frac{m^{2}}{z^{4}g}\right]F=0.
\end{eqnarray}
From the Sturm-Liouville eigenvalue problem \cite{Gelfand-Fomin}, we
write down the expression which can be used to estimate the minimum
eigenvalue of $\lambda^2$
\begin{eqnarray}\label{SLEigenvalue}
\lambda^{2}=\frac{\int^{1}_{0}T\left(F'^{2}-UF^{2}\right)dz}{\int^{1}_{0}TVF^{2}dz},
\end{eqnarray}
with
\begin{eqnarray}
&&U=\frac{\Delta_{i}}{z}\left(\frac{\Delta_{i}+3-d}{z}+\frac{g'}{g}\right)-\frac{m^{2}}{z^{4}g},\nonumber\\
&&V=\frac{(1-z^{d-3})^{2}}{z^{4}g^{2}}.
\end{eqnarray}
In order to use the variation method, we will assume the trial
function to be $F(z)=1-az^{2}$, where $a$ is a constant.

Using Eq. (\ref{SLEigenvalue}) to compute the minimum eigenvalue of
$\lambda^{2}$ for $i=+$ or $i=-$, we can obtain the critical
temperature $T_{c}$ for different strength of the backreaction
$\kappa$ and the mass of the scalar field $m$ from the following
relation
\begin{eqnarray}\label{CTTc}
T_{c}=\frac{1}{4\pi}\left[(d-1)-\frac{(d-3)^{2}}{d-2}\kappa_{n}^{2}(\lambda^{2}|_{\kappa_{n-1}})\right]
\left(\frac{\rho}{\lambda}\right)^{\frac{1}{d-2}}.
\end{eqnarray}
As an example, we calculate the case for $d=5$ and $m^{2}L^2=-3$
with the chosen values of the backreaction parameter $\kappa$ for
$i=+$, i.e., $\Delta_{+}=3$. Setting $\Delta\kappa=0.05$, for
$\kappa_{0}=0$ we have
\begin{eqnarray}
\lambda^{2}=\frac{2(-18+27a-14a^{2})}{6(4\ln2-3)+16(3\ln2-2)a+(24\ln2-17)a^{2}},
\end{eqnarray}
whose minimum is $\lambda^{2}|_{\kappa_{0}}=18.23$ at $a=0.7218$.
According to Eq. (\ref{CTTc}), we can easily get the critical
temperature $T_{c}=0.1962\rho^{1/3}$, which agrees well with the
numerical result $T_{c}=0.1980\rho^{1/3}$ \cite{HorowitzPRD78}. For
$\kappa_{1}=0.05$, substituting $\lambda^{2}|_{\kappa_{0}}$ into
Eqs. (\ref{gzNew}) and  (\ref{TzNew}) we obtain
\begin{eqnarray}
\lambda^{2}=\frac{4.466-6.682a+3.462a^{2}}{0.1714-0.1600a+0.04592a^{2}},
\end{eqnarray}
which attains its minimum $\lambda^{2}|_{\kappa_{1}}=18.11$ at
$a=0.7195$. Hence the critical temperature reads
$T_{c}=0.1934\rho^{1/3}$, which is also in good agreement with the
numerical result $T_{c}=0.1953\rho^{1/3}$. For $\kappa_{2}=0.10$,
putting $\lambda^{2}|_{\kappa_{1}}$ in Eqs. (\ref{gzNew}) and
(\ref{TzNew}) we arrive at
\begin{eqnarray}
\lambda^{2}=\frac{4.364-6.478a+3.349a^{2}}{0.1740-0.1633a+0.04705a^{2}},
\end{eqnarray}
whose minimum is $\lambda^{2}|_{\kappa_{2}}=17.75$ at $a=0.7122$. So
the critical temperature is $T_{c}=0.1852\rho^{1/3}$, which is again
consistent with the numerical finding $T_{c}=0.1874\rho^{1/3}$. For
other values of $\kappa$, the similar iterative procedure also can
be applied to present the analytic result for the critical
temperature. When we reduce the  step size, for example to fix
$\Delta\kappa=0.025$, we can also compute the critical temperature
$T_{c}$ in the similar way. In Table \ref{CriticalTcD5} we give the
critical temperature $T_{c}$ for the scalar operator
$<\mathcal{O}_{+}>$ when we fix the mass of the scalar field
$m^{2}L^2=-3$ for different strength of the backreaction by choosing
the step size $\Delta\kappa=0.05$ and $0.025$, respectively. We find
that the analytic results derived from the S-L method are in very
good agreement with the numerical calculation. Furthermore we
observe that when we reduce the step size $\Delta\kappa$, we can
improve the analytic result and get the critical temperature more
consistent with the numerical result.

\begin{table}[ht]
\caption{\label{CriticalTcD5} The critical temperature $T_{c}$ with
the chosen values of the backreaction parameter $\kappa$ and the
step size $\Delta\kappa$ for the condensates of the scalar operator
$<\mathcal{O}_{+}>$ in the case of 5-dimensional AdS black hole
background. Here we fix the mass of the scalar field by
$m^{2}L^2=-3$.}
\begin{tabular}{c c c c}
         \hline
~ & Analytical($\Delta\kappa=0.05$)~~~~&~~~~
Analytical($\Delta\kappa=0.025$) & Numerical
        \\
        \hline
~~~~$\kappa=0$~~~~~~~~&~~~~~~~~$0.1962\rho^{1/3}$~~~~~~~~&~~~~~~~~$0.1962\rho^{1/3}$~~~~~~~~&~~~~~~~~$0.1980\rho^{1/3}$~~~~~~~~
          \\
~~~~$\kappa=0.05$~~~~&~~~~~~~~$0.1934\rho^{1/3}$~~~~~~~~&~~~~~~~~$0.1934\rho^{1/3}$~~~~~~~~&~~~~~~~~$0.1953\rho^{1/3}$~~~~~~~~
          \\
~~~~$\kappa=0.10$~~~~&~~~~~~~~$0.1852\rho^{1/3}$~~~~~~~~&~~~~~~~~$0.1853\rho^{1/3}$~~~~~~~~&~~~~~~~~$0.1874\rho^{1/3}$~~~~~~~~
          \\
~~~~$\kappa=0.15$~~~~&~~~~~~~~$0.1718\rho^{1/3}$~~~~~~~~&~~~~~~~~$0.1722\rho^{1/3}$~~~~~~~~&~~~~~~~~$0.1748\rho^{1/3}$~~~~~~~~
          \\
~~~~$\kappa=0.20$~~~~&~~~~~~~~$0.1540\rho^{1/3}$~~~~~~~~&~~~~~~~~$0.1549\rho^{1/3}$~~~~~~~~&~~~~~~~~$0.1580\rho^{1/3}$~~~~~~~~
          \\
~~~~$\kappa=0.25$~~~~&~~~~~~~~$0.1330\rho^{1/3}$~~~~~~~~&~~~~~~~~$0.1345\rho^{1/3}$~~~~~~~~&~~~~~~~~$0.1382\rho^{1/3}$~~~~~~~~
          \\
~~~~$\kappa=0.30$~~~~&~~~~~~~~$0.1098\rho^{1/3}$~~~~~~~~&~~~~~~~~$0.1123\rho^{1/3}$~~~~~~~~&~~~~~~~~$0.1165\rho^{1/3}$~~~~~~~~
          \\
        \hline
\end{tabular}
\end{table}

For completeness, we also extend the investigation to the
4-dimensional AdS black hole background. In Table
\ref{CriticalTcD4}, we present the critical temperature $T_{c}$ of
the chosen parameter $\kappa$ with the scalar operators
$<\mathcal{O}_{-}>$ and $<\mathcal{O}_{+}>$ for the
($2+1$)-dimensional superconductor if we fix the mass of the scalar
field by $m^{2}L^2=-2$ and the step size by $\Delta\kappa=0.05$. The
agreement of the analytic results derived from S-L method with the
numerical calculation shown in Tables \ref{CriticalTcD5} and
\ref{CriticalTcD4} is impressive.

\begin{table}[ht]
\caption{\label{CriticalTcD4} The critical temperature $T_{c}$
obtained by the analytical S-L method (left column) and from
numerical calculation (right column) with the chosen values of the
backreaction parameter $\kappa$ for the condensates of the scalar
operators $<\mathcal{O}_{-}>$ and $<\mathcal{O}_{+}>$ in the case of
4-dimensional AdS black hole background. Here we fix the mass of the
scalar field by $m^{2}L^2=-2$ and the step size by
$\Delta\kappa=0.05$.}
\begin{tabular}{c c c}
         \hline
~ & $<\mathcal{O}_{-}>$ & $<\mathcal{O}_{+}>$
        \\
        \hline
~~~~$\kappa=0$~~~~~~~~&~~~~~~~~$0.2250\rho^{1/2}$~~~~~~~$0.2255\rho^{1/2}$~~~~~~~~&~~~~~~~~$0.1170\rho^{1/2}$~~~~~~~$0.1184\rho^{1/2}$~~~~~
          \\
~~~~$\kappa=0.05$~~~~&~~~~~~~~$0.2249\rho^{1/2}$~~~~~~~$0.2253\rho^{1/2}$~~~~~~~~&~~~~~~~~$0.1163\rho^{1/2}$~~~~~~~$0.1177\rho^{1/2}$~~~~~
          \\
~~~~$\kappa=0.10$~~~~&~~~~~~~~$0.2246\rho^{1/2}$~~~~~~~$0.2250\rho^{1/2}$~~~~~~~~&~~~~~~~~$0.1141\rho^{1/2}$~~~~~~~$0.1156\rho^{1/2}$~~~~~
          \\
~~~~$\kappa=0.15$~~~~&~~~~~~~~$0.2241\rho^{1/2}$~~~~~~~$0.2245\rho^{1/2}$~~~~~~~~&~~~~~~~~$0.1106\rho^{1/2}$~~~~~~~$0.1121\rho^{1/2}$~~~~~
          \\
~~~~$\kappa=0.20$~~~~&~~~~~~~~$0.2235\rho^{1/2}$~~~~~~~$0.2239\rho^{1/2}$~~~~~~~~&~~~~~~~~$0.1057\rho^{1/2}$~~~~~~~$0.1074\rho^{1/2}$~~~~~
          \\
~~~~$\kappa=0.25$~~~~&~~~~~~~~$0.2226\rho^{1/2}$~~~~~~~$0.2230\rho^{1/2}$~~~~~~~~&~~~~~~~~$0.0998\rho^{1/2}$~~~~~~~$0.1017\rho^{1/2}$~~~~~
          \\
~~~~$\kappa=0.30$~~~~&~~~~~~~~$0.2216\rho^{1/2}$~~~~~~~$0.2220\rho^{1/2}$~~~~~~~~&~~~~~~~~$0.0929\rho^{1/2}$~~~~~~~$0.0951\rho^{1/2}$~~~~~
          \\
        \hline
\end{tabular}
\end{table}
It further supports the observation obtained first in the numerical
computation that the stronger backreaction can make the scalar hair
more difficult to be developed
\cite{Gubser-Nellore,Aprile-Russo,Brihaye,Liu-Sun,BarclayGregory,Siani,
Barclay2011,PanWangBR,LiuWangBTZ,GregoryRev,LiuBIBR,HGST2}. The
consistency between the analytic and numerical results indicates
that the S-L method is a powerful analytic way to investigate the
holographic superconductor even when we take the backreaction  into
account.

\section{Conclusions}

We have generalized the variational method for the Sturm-Liouville
eigenvalue problem to analytically investigate the properties of the
holographic superconductor  with backreactions. We found that in the
fully backreacted spacetime, the S-L method is still powerful to
disclose the property of the condensation. Our analytic results are
in very good agreement with those obtained from numerical
computations. If we reduce the step in the iterative procedure, we
can further improve our analytic results and improve the consistency
with the numerical findings. Our analytic result shows that the
backreaction makes the critical temperature of the superconductor
decrease, which can be used to back up the numerical finding as shown in figure 2 of Ref. \cite{HartnollJHEP12} that the backreaction can hinder the condensation to be formed.

\begin{acknowledgments}

This work was supported by the National Natural Science Foundation
of China; the National Basic Research of China under Grant No.
2010CB833004, PCSIRT under Grant No. IRT0964, NCET under Grant No.
10-0165, the Construct Program of the National Key Discipline, and
Hunan Provincial Natural Science Foundation of China 11JJ7001.
\end{acknowledgments}

\end{document}